\documentclass{PoS}

\usepackage{epsfig,multicol}
\usepackage[final]{showkeys}
\usepackage{amsmath}
\usepackage{amssymb}

\newcommand{\bel}{\begin{equation}\label}

\newcommand {\beq}{\begin{equation}}
\newcommand {\eeq}{\end{equation}}
\newcommand {\beqa}{\begin{eqnarray}}
\newcommand {\eeqa}{\end{eqnarray}}
\newcommand {\bc}{\begin{center}}
\newcommand {\ec}{\end{center}}
\newcommand {\tr}{{\rm tr\,}}

\newcommand {\vev}  [1]{\ensuremath{\langle #1 \rangle}}
\newcommand {\rf}   [1]{(\ref{#1})}

\def\vs5{\vspace*{5mm}}
\def\vs1{\vspace*{1cm}}
\def\vs2{\vspace*{2cm}}
\def\hs5{\vspace*{5mm}}
\def\hs1{\hspace*{1cm}}
\def\hs2{\hspace*{2cm}}
\def\vs50{\vspace*{50mm}}
\def\vs20{\vspace*{20mm}}

\def\tr{\hbox{tr}}

\newcommand\jhep [3]{{{\it J. High Energy Phys.\ }{\bf #1} (#2) #3}}

\newcommand\ptp  [3]{{{\it Prog.\ Theor.\ Phys.\ }{\bf #1} (#2) #3}}

\title{Monte Carlo simulations of a supersymmetric matrix model of dynamical compactification in non perturbative string theory}

\ShortTitle{SUSY matrix model of dynamical compactification}

\author{\speaker{Konstantinos N. Anagnostopoulos}%
         \thanks{The work of K.N.A.\ was partially funded 
by the National Technical University of
Athens through the Basic Research Support Programmes 2009 and 2010.
The work of T.A. and J.N.\ is supported in part by Grant-in-Aid 
for Scientific Research 
(No.\ 23740211 for T.A. and 20540286, 23244057 for J.N.)
from Japan Society for the Promotion of Science.}\\
        Physics Department, 
        National Technical University of Athens\\ 
        Zografou Campus, 157--80 Zografou, Greece\\
        E-mail: \email{konstant@mail.ntua.gr}}

\author{Takehiro Azuma\\
        Institute for Fundamental Sciences, Setsunan University,
        17-8 Ikeda Nakamachi, Neyagawa, Osaka 572-8508, Japan\\
        E-mail: \email{azuma@mpg.setsunan.ac.jp}}

\author{Jun Nishimura\\
        High Energy Accelerator Research Organization (KEK) and
        Graduate University for Advanced Studies (SOKENDAI),  
        1-1 Oho, Tsukuba 305-0801, Japan\\
        E-mail: \email{jnishi@post.kek.jp}}

\abstract{%
The IKKT or IIB matrix model has been postulated to be a non
perturbative definition of superstring theory. It has the attractive
feature that spacetime is dynamically generated, which makes possible
the scenario of dynamical compactification of extra dimensions, which
in the Euclidean model manifests by spontaneously breaking the SO(10)
rotational invariance (SSB). In this work we study using Monte Carlo
simulations the 6 dimensional version of the Euclidean IIB matrix
model.  Simulations are found to be plagued by a strong complex action
problem and the factorization method is used for effective sampling
and computing expectation values of the extent of spacetime in various
dimensions.  Our results are consistent with calculations using the
Gaussian Expansion method which predict SSB to SO(3) symmetric vacua,
a finite universal extent of the compactified dimensions and  finite
spacetime volume.}

\FullConference{The 30th International Symposium on Lattice Field Theory\\
                 June 24 – 29,  2012\\
                 Cairns, Australia}

\begin{document}

\section{Introduction}
Superstring theory is a natural candidate of a unified theory of all
interactions, including gravity. It lacks a non-perturbative
definition, which would allow us to address dynamically questions such
as the preferred vacuum where our universe sits and determine its
properties such as classical spacetime dimensionality. The large $N$
limit of the 10 dimensional IKKT or IIB matrix model has been proposed
to provide such a definition \cite{Ishibashi:1996xs}.  Using
dualities, which suggest that the five types of superstring theory are
actually limits of a unique theory, the IIB matrix model is expected
to describe the unique underlying theory despite its explicit
connection to perturbative type IIB superstring theory.

The IIB matrix model has a series of attractive features: The model
has a unique scale, which raises the possibility for the theory to
choose a unique vacuum. Spacetime and matter content can arise
dynamically from the distribution of the eigenvalues of the bosonic
matrices, which makes possible the realization of the scenario of
dynamical compactification of extra dimensions. The motivation for
studying such a scenario in the Euclidean model comes from noticing
that lower dimensional configurations are stationary points of the
fluctuating complex part of the action of the model
\cite{Nishimura:2000ds}. Using the 
Gaussian Expansion Method (GEM), one can show that the SO(3) symmetric
vacuum has the smallest free energy density compared to higher or
lower dimensional configurations \cite{Nishimura:2001sx}. Dynamical
compactification, 
therefore, occurs by spontaneously breaking the SO(10) rotational
symmetry of the model (SSB).  Moreover, a universal scale $r$ for the
small dimensions is computed\footnote{Note that the actual length of
  each dimension is, by definition, $r^{1/2}$ and $R^{1/2}$
  respectively.}, with the scale $R$ of the large dimensions being
such that $R^d r^{10-d} = \ell^{10}$, with $\ell$ being the scale of
the size of the symmetric configurations. The former relation is
interpreted as a constant volume property of the $d$ dimensional
(metastable) vacua of the model and leads to the conclusion that
spacetime in the Euclidean IIB model has finite volume
\cite{Nishimura:2001sx}. 

Monte Carlo simulations of the Euclidean IIB matrix model, as well as
simpler related models, can confirm these results from first
principles and shine light into the mechanism of dynamical
compactification of extra dimensions
\cite{Krauth:1998xh,Ambjorn:2000bf}. Such simulations are hindered by
a severe complex action problem and are quite difficult. Some results,
however, have been obtained by studying related lower dimensional toy
models using the so called factorization method. This method, proposed
originally in \cite{0108041} has been tested also in Random Matrix
theory and finite density QCD \cite{Ambjorn:2003rr}. It attempts to
sample the most important configurations that contribute to the
partition function, first by numerically solving equations that
compute the maxima of the distribution functions of appropriately
chosen observables that are strongly correlated to the fluctuating
phase \cite{multifac} and then by simulating a constrained system in
the region of the solutions. This makes importance sampling possible by
also taking into account the suppression of configurations by the
fluctuations of the phase, together with the suppression caused by the
measure and the real part of the action in the partition function. The
overlap problem is thus solved and the complex action problem reduced
due to milder fluctuations of the phase within the sampled
configurations and the use of scaling properties that permit
extrapolations to larger systems.

In this work, we present Monte Carlo simulation results of a 6
dimensional version of the IIB Matrix Model. Contrary to
\cite{0108041}, where the oneloop approximation was used, the full
model is simulated. This is necessary since, as we will show, the
short distance non perturbative dynamics of the eigenvalues of the
matrices play a crucial role in generating the scales of dynamical
compactification. We show that in the absence of the complex part of
the action, no SSB of the SO(6) rotational invariance occurs. We apply
the factorization method in order to compute the eigenvalues
$\lambda_1>\lambda_2>\ldots>\lambda_6$ of spacetime's ``moment of
inertia'' tensor $T_{\mu\nu}$ and compare our results with those
obtained in \cite{Nishimura:2001sx} using the GEM. In particular, we
test the GEM predictions of SSB to SO(3) symmetric vacua, a universal
scale $r$ for compactified dimensions and finite
spacetime volume. Our results are consistent with these scenaria.

\section{The Model}
We study a 6 dimensional version of the IIB matrix model defined by
the partition function
\begin{equation}
\label{2.1}
Z = \int dA\,d\psi\,d\bar\psi\, {\rm e}^{-S_b-S_f}\, ,
\end{equation}
where
$S_b = -\dfrac{N}{4}\tr \left[ A_\mu,A_\nu\right]^2$ and
$S_f =  \dfrac{N}{2}\tr\bar\psi_\alpha(\Gamma)_{\alpha\beta}\left[A_\mu,\psi_\beta\right]$.
The model can be formally obtained by the dimensional reduction of the
Euclidean 6 dimensional ${\cal N}=1$ U($N$) Super Yang Mills theory
to zero dimensions: The $N\times N$ matrices $A_\mu$, $\mu=1,\ldots,6$
are traceless and hermitian and transform as O(6) vectors. The
$N\times N$ matrices $\psi_\alpha$, $\bar\psi_\alpha$ are traceless
with grassmannian entries and transform as Weyl spinors. The model
turns out to have ${\cal N}=2$ supersymmetry, which leads to the
interpretation of the eigenvalues of $A_\mu$ defining the points of
the 6 dimensional (Euclidean) spacetime
\footnote{Albeit a non 
  classical spacetime with a fuzzy geometry since the dominant
  configurations of $A_\mu$ cannot be simultaneously diagonalizable
  \cite{Ambjorn:2000bf}.} \cite{Ishibashi:1996xs}. Dynamical
compactification can manifest if the 
distribution of these points break O(6) rotational invariance
spontaneously (SSB). The order parameters of such SSB are the
expectation values $\vev{\lambda_1},\ldots,\vev{\lambda_6}$ of the
``moment of inertia tensor'' of spacetime
$T_{\mu\nu}=\dfrac{1}{N}\tr\, A_\mu A_\nu$,
where the eigenvalues $\lambda_1>\ldots>\lambda_6$ are ordered {\it
  before} taking the expectation value. SSB occurs if, in the large
$N$ limit, some of the expectation values
$\vev{\lambda_1},\ldots,\vev{\lambda_d}$ grow ``large'' and the
remaining $\vev{\lambda_{d+1}},\ldots,\vev{\lambda_6}$ remain
small. The large eigenvalues define the extended dimensions of
spacetime and we obtain {\it dynamically} a $d$ dimensional
spacetime. This scenario has been studied using GEM in
\cite{Nishimura:2001sx} where 
SO(3) invariant configurations were found to have the minimum free
energy density, thus dominating the path integral \rf{2.1}. The $6-d$
small 
dimensions in the SO(d) vacua turn out to have a
$d$--independent thickness $r^{1/2}$, whereas the $d$ large ones have
thickness $R^{1/2}$, such that
\begin{equation}
\label{2.4}
R^d \, r^{6-d} = \ell^6\, ,
\end{equation}
where $\ell\approx 0.627$ is the square of the extent of the SO(6)
symmetric configurations, $\vev{\lambda_1}=\ldots=\vev{\lambda_d}=R$
and $\vev{\lambda_{d+1}}=\ldots=\vev{\lambda_6}=r$. The SO(6)
symmetric configurations dominate in the {\it phase quenched model}
discussed below where 
SSB does not occur \cite{Ambjorn:2000bf}. Eq. \rf{2.4} expresses the
constant volume 
property of the SO($d$) vacua and implies that the volume of spacetime
in the 6 dimensional Euclidean IIB matrix model is {\it finite}
\cite{Nishimura:2001sx}. 

Monte Carlo simulations are performed by integrating out the fermions
in \rf{2.1} first, obtaining
\begin{equation}
\label{2.5}
Z= \int\, dA\, \det{\cal M}\, {\rm e}^{-S_b} =
 \int\, dA\, {\rm e}^{-S_0+ i \Gamma} \, ,
\end{equation}
where $\det{\cal M} = $ $Z_f[A] = $  
$ \displaystyle\int d\psi\,d\bar\psi {\rm
  e}^{-S_f}$ and $S_0=S_b-\log|\det{\cal M}|$. The determinant $\det{\cal M}=|\det{\cal M}|{\rm
  e}^{i\Gamma}$ turns out to be 
generically complex creating a {\it very strong} complex action
problem in the simulations. We applied the factorization method
\cite{0108041}, which first amounts to considering the phase quenched
model $\displaystyle Z_0 = \int\, dA\, {\rm e}^{-S_0}$.
In this model, we computed the
phase quenched expectation values
$\vev{\lambda_1}_0,\ldots,\vev{\lambda_n}_0$, and defined the
normalized eigenvalues
$\tilde\lambda_n = {\lambda_n}/{\vev{\lambda_n}_0}$.
Deviation of $\vev{\tilde\lambda_n}$ from 1, is the result of the
suppression of the dominant configurations of $Z_0$ by the fluctuations
of the 
phase $\Gamma$. Due to the strong correlations of $\lambda_n$ with
$\Gamma$, we consider the distribution functions \cite{0108041,multifac}
\begin{equation}
\label{2.8}
\rho_n(x)=\vev{\delta(x-\tilde\lambda_n)} =
 \frac{1}{C}\rho^{(0)}_n(x) w_n(x)\, ,
\end{equation}
where\footnote{The constant $C=\vev{{\rm e}^{i\Gamma}}_0$ is not
  necessary in the calculations described below.}
$\rho^{(0)}_n(x)=\vev{\delta(x-\tilde\lambda_n)}_0$, $w_n(x)=\vev{{\rm
    e}^{i\Gamma}}_{n,x}$ and $\vev{\cdot}_{n,x}$ denotes expectation
values within the constrained system
$\displaystyle Z_{n,x} = \int\, dA\, {\rm e}^{-S_0}\,\delta(x-\tilde\lambda_n)$.

The value of $\vev{\lambda_n}$ in the large $N$ limit
is determined by the minimum of the
``free energy'' ${\cal F}_n(x)=-\log\rho_n(x)$ which,  given
Eq. \rf{2.8}, is the large $N$ limit of a solution to
\begin{equation}
\label{2.10}
 \frac{1}{N^2}f^{(0)}_n(x)\equiv 
 \frac{1}{N^2}\frac{d}{dx}\log\rho^{(0)}_n(x)=
-\frac{1}{N^2}\frac{d}{dx}\log w_n(x)
\, .
\end{equation}
Using the dominant 
solution of \rf{2.10} as an estimator of $\vev{\lambda_n}$ for
finite $N$, has the advantages that the overlap problem is solved by
simulating $Z_{n,x}$ and the complex action problem reduced, since
$\Phi_n(x)=\lim\limits_{N\to\infty} \dfrac{1}{N^2}\log w_n(x)$ is found to
scale for relatively small $N$ and it is possible to extrapolate
solutions to larger values of $N$. Moreover, the numerical errors of
the solution do not grow exponentially with $N$
\cite{0108041,Ambjorn:2003rr}. 
\section{Simulations and Results}
Monte Carlo simulations are performed on the system
$\displaystyle Z_{n,V}=\int\, dA\, {\rm e}^{-S_0-V(\lambda_n)}$ where $V(z)=\frac{1}{2}\gamma(z-\xi)^2$
and $\gamma$, $\xi$ are parameters. The rational hybrid Monte
Carlo method (RHMC) is used in the simulations. We take $\gamma$ large
enough, so that $\rho_{n,V}(x)$ is sharply peaked
at $x_p$ and the results are independent of $\gamma$. We use the
estimators $x_p=\vev{\tilde\lambda_n}_{n,V}$, $w_n(x_p)=\vev{{\rm
    e}^{i\Gamma}}_{n,V}$ and
  $f^{(0)}_n(x_p)=\vev{\lambda_n}_0V'(\vev{\lambda_n}_{n,V})$. 

\begin{figure}
  \begin{center}
    \epsfig{file=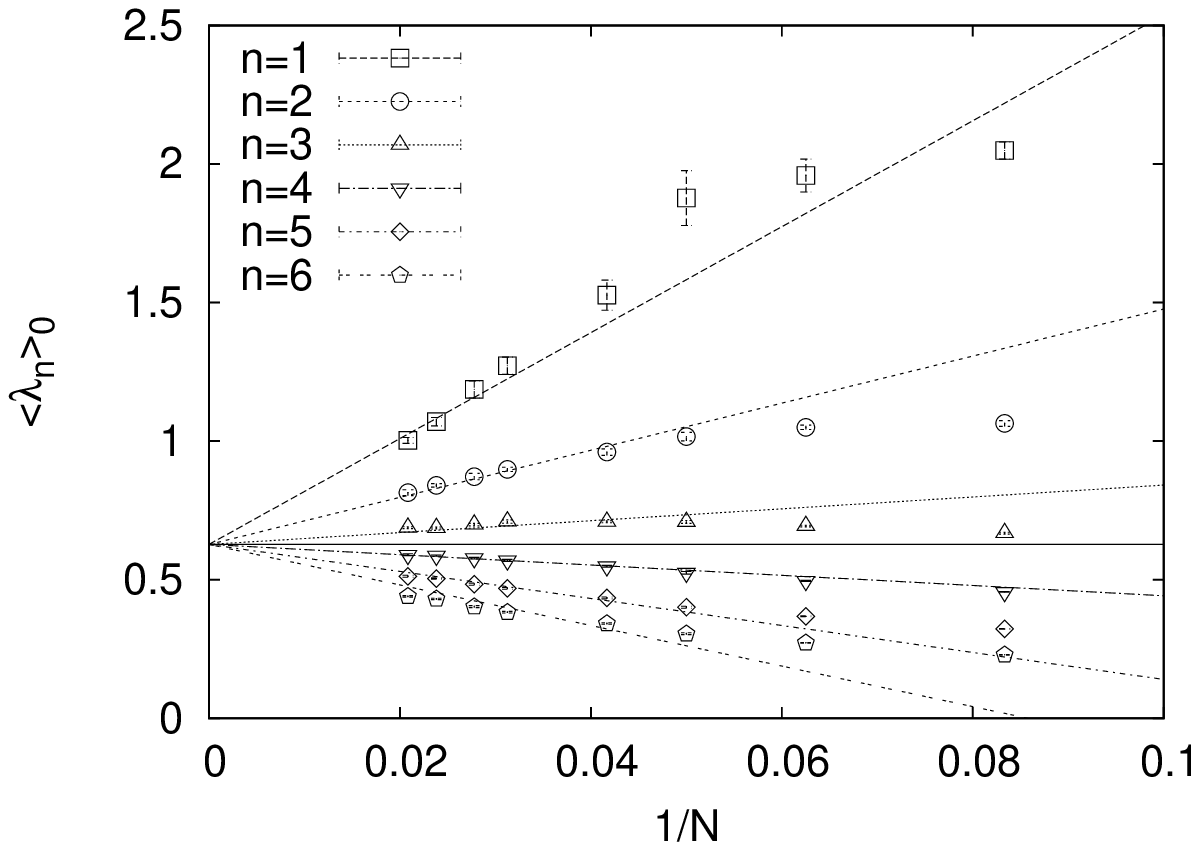    ,width=7.4cm}
    \epsfig{file=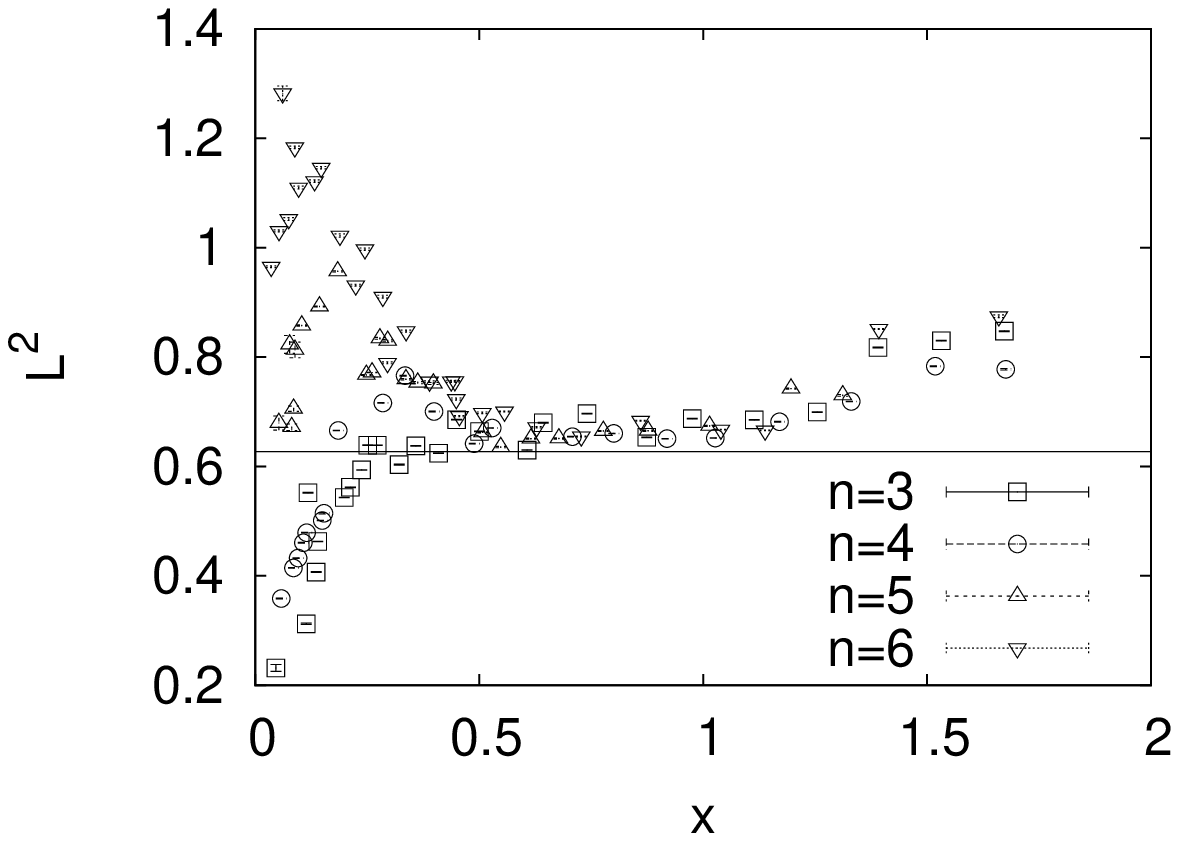,width=7.4cm}
  \end{center}
    \caption{(LEFT) The $\langle \lambda_{n} \rangle_{0}$ of the phase
      quenched model are plotted against
      $1/N$ for $12\le N\le 48$. The horizontal line is $\ell = 0.627$.
    (RIGHT) $L_n^2 = \left(\prod\limits_{i=1}^6 \vev{\lambda_i}_{n,V}\right)^{1/6}$ for $N=32$.
  The horizontal line is $\ell = 0.627$.} 
   \label{f:01}
\end{figure}

For $\gamma=0$ we obtain the phase quenched model $Z_0$. We simulate
this system and show the results for $\langle \lambda_{n}
\rangle_{0}$, $n=1,\ldots,6$ in the left plot of Fig. \ref{f:01}. In
the large $N$ limit, these values converge to the same value
$\ell\approx 0.627$ as predicted by GEM \cite{Nishimura:2001sx}. No
SSB of SO(6) occurs in the absence of a fluctuating phase $\Gamma$ as
expected \cite{Nishimura:2000ds,Ambjorn:2000bf}.

\begin{figure}
  \begin{center}
    \epsfig{file=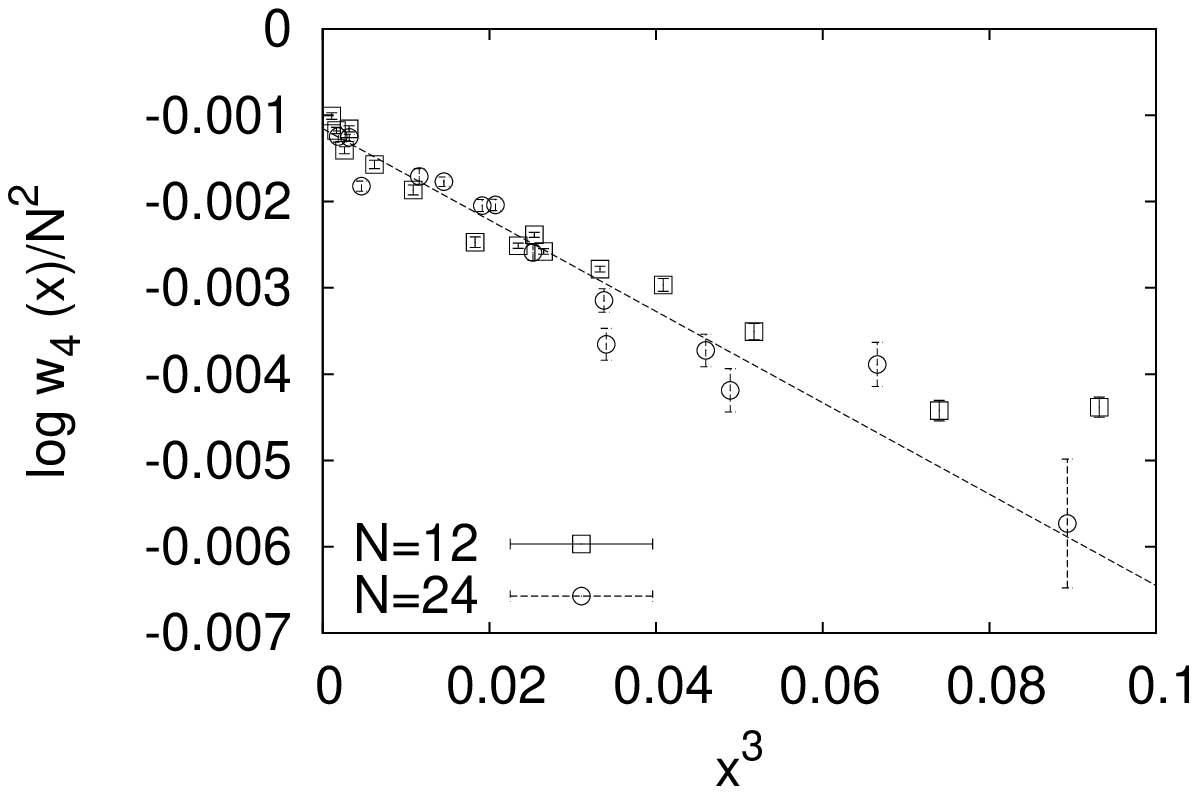,width=7.4cm}
    \epsfig{file=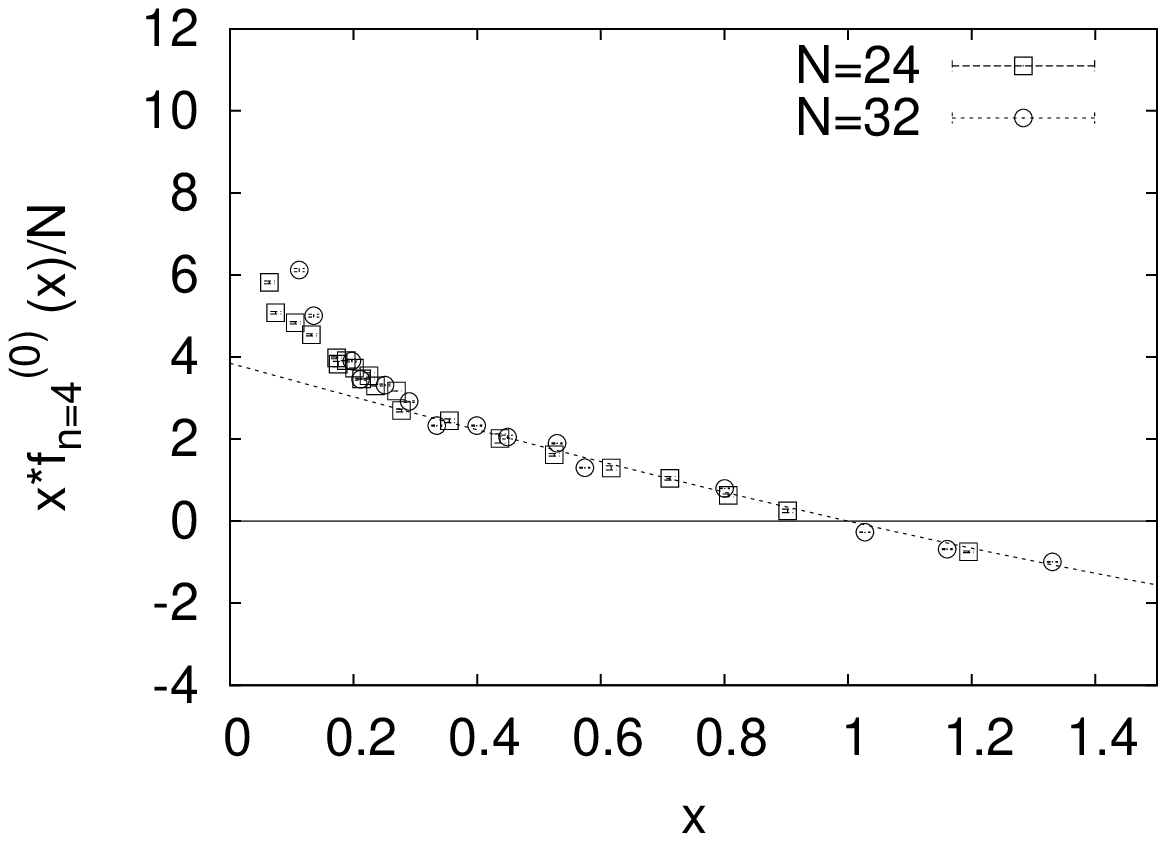   ,width=7.4cm}
  \end{center}
    \caption{(LEFT) Small $x$ scaling of  $\log w_4(x)/N^2$,
      according to the discussion on page \protect\pageref{p:01} for $N=24$. 
  (RIGHT) $\dfrac{1}{N} f^{(0)}_4(x)$ 
      scaling around the peak $x\approx 1$,
      according to the discussion on page \protect\pageref{p:02} for
      $N=24$. The fit is the function $g_4(x)$ described in the text.} 
   \label{f:02}
\end{figure}

\begin{figure}
  \begin{center}
    \epsfig{file=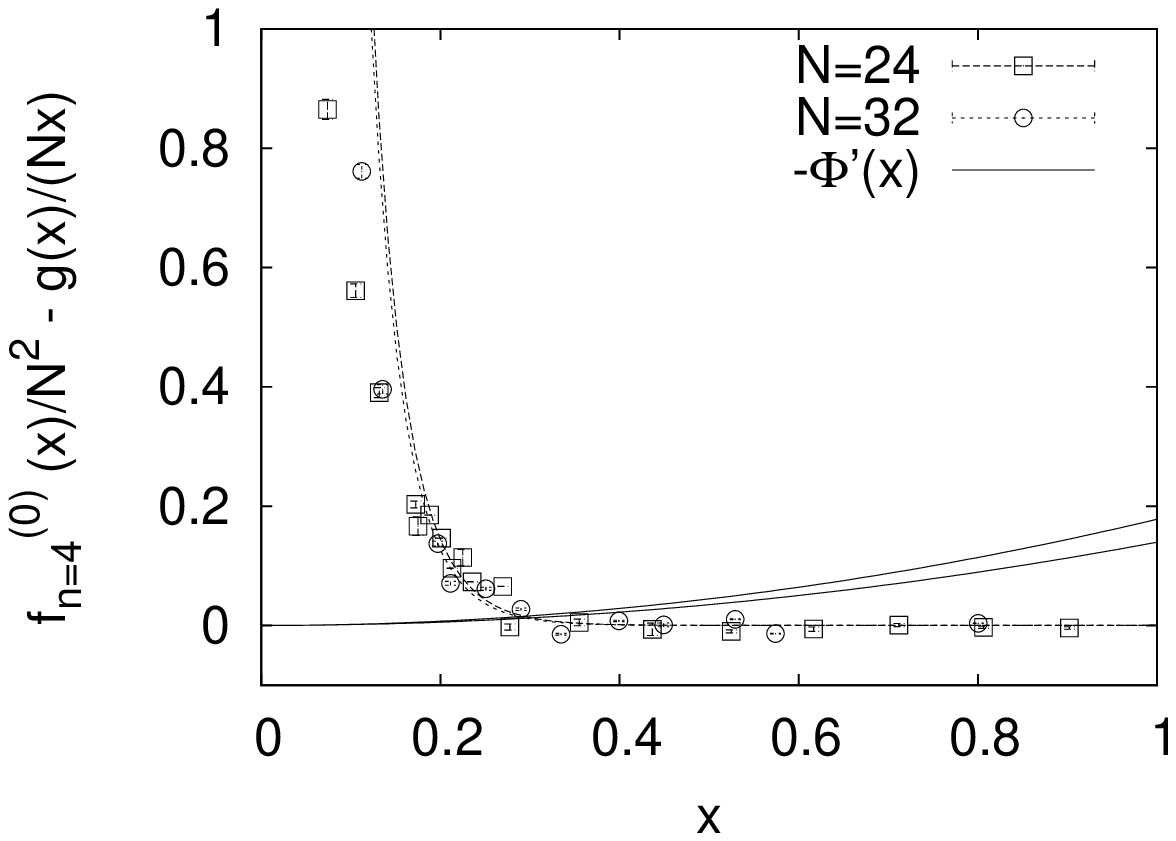         ,width=7.4cm}
    \epsfig{file=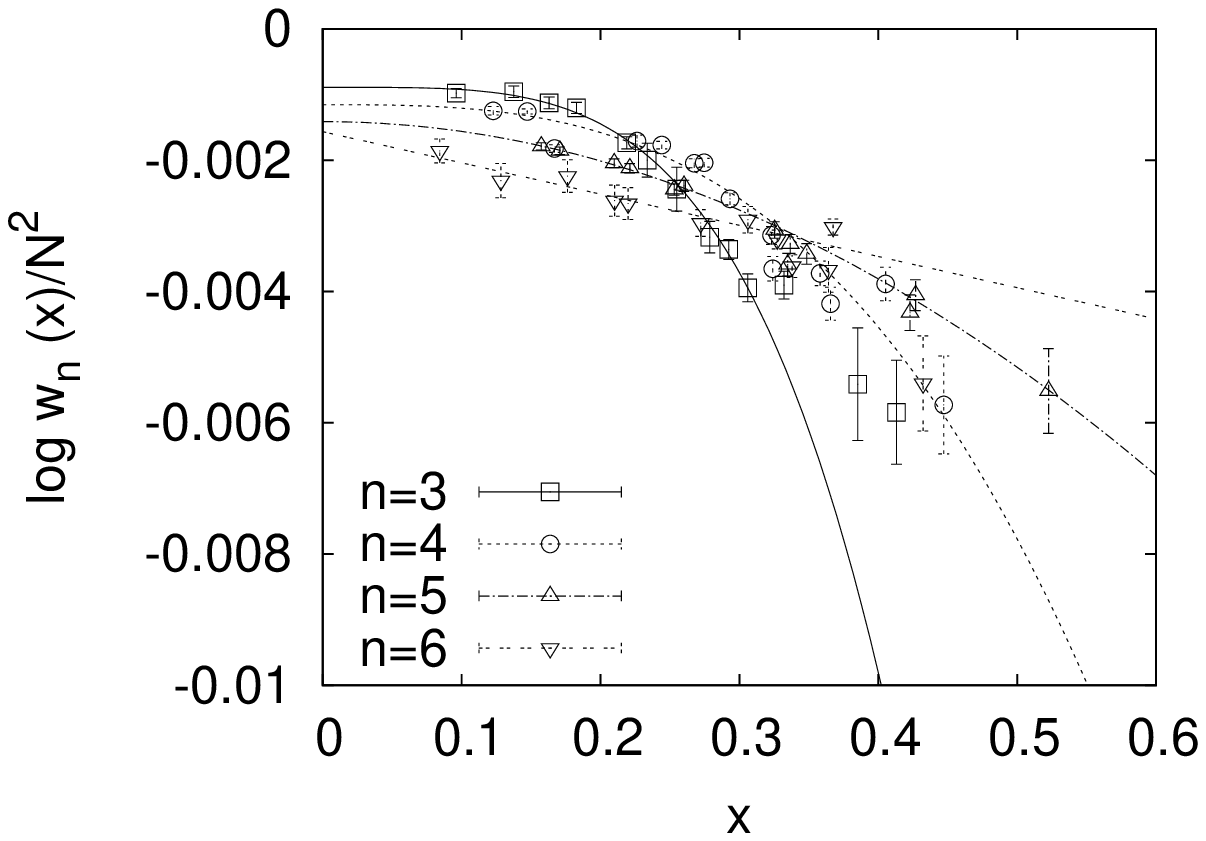  ,width=7.4cm}
  \end{center}
    \caption{(LEFT) The solution of Eq. \protect\rf{2.10}, giving
      $\vev{\tilde\lambda_4}=0.31(2)$. The vertical axis and the
      corresponding fit of the data points estimate the
      LHS of \protect\rf{2.10} in the large $N$ limit. 
      (RIGHT) $\log w_n(x)/N^2$ 
      for $N=24$ with the
      corresponding fits to their small $x$ scaling. These curves can
      be used to compute the difference $\Delta_{34}$ of Eq. \protect\rf{3.2}.} 
   \label{f:03}
\end{figure}

In order to compute the solution of \rf{2.10} in the large $N$ limit,
it is important to use the scaling properties of the functions
$w_n(x)$ and $f^{(0)}_n(x)$ for small $x$ and in the large $N$ limit.
We find that
$\Phi_n(x)$, $n=3, 4, 5, 6$ is almost constant for large $x$. Thus the
extended direction becomes decorrelated with the phase and there is no
need to constrain the large eigenvalues as in \cite{multifac}. We also
observe good scaling of $\dfrac{1}{N^2}\log w_n(x)$ for $12<N<24$. For
solving \rf{2.10}, the small $x$ scaling behavior \cite{multifac}
$\dfrac{1}{N^2}\log w_n(x)\sim -c_{0,n} - c_{1,n} x^{7-n}$,
$n=2,\ldots,6$ is important. We compute the coefficients $c_{0,n}$ and
$c_{1,n}$, as shown in the left plot of Fig. \ref{f:02}. For
$f^{(0)}_n(x)$ the expected small $x$ scaling is
$\dfrac{1}{N^2}f^{(0)}_n(x)\sim \left\{\dfrac{1}{2}(7-n)+2\delta_{n
  1}\right\}\dfrac{1}{x}$ and implies the existence of a hard core
potential suppressing the small $x$ region, as it was first found in
\cite{Ambjorn:2000bf}. This is a strictly non perturbative effect
which is absent e.g. in the one loop
approximation\cite{0108041}. Around $x=1$ a $\dfrac{1}{N}
f^{(0)}_n(x)$ \label{p:02} scaling is found to dominate\footnote{Note
  that in the one loop approximation, the $\dfrac{1}{N} f^{(0)}_n(x)$
  scaling holds {\it for all} values of $x$ \cite{0108041}.}.  This
adds finite size effects to the small--$x$ scaling which can be
subtracted by removing the ${\cal O}(1/N)$ term $g_n(x)/N x$, where we
take $g_n(x)=a_{1,n}(x-1)+a_{2,n}(x-1)^2+a_{3,n}(x-1)^3$ around the peak
$x\approx 1$. The computation of $g_n(x)$ is done by the fit in the
right plot of Fig. \ref{f:02}, where we show the result for
$n=4$. Finally, the left hand side of \rf{2.10} is estimated by the
function\label{p:01} $\dfrac{1}{N^2}f^{(0)}_n(x)-\dfrac{g_n(x)}{N x} =
$ $\left\{\dfrac{1}{2}(7-n)+2\delta_{n 1}\right\}\dfrac{\exp{(-q_n
    x)}}{x}$, where $q_n$ is determined by the fit of the left plot of
Fig. \ref{f:03}. In the same plot we also show $-\Phi'_n(x)$, and the
intersection of the two curves estimates the solution of \rf{2.10} for
$n=4$. It is found that $\vev{\tilde\lambda_4} = 0.31(2)$. The
corresponding GEM prediction is $r/\ell\approx 0.223/0.627 = 0.355$
\cite{Nishimura:2001sx}. Preliminary results for $n=3,5$ and $6$ yield
similar results which are consistent with the GEM finding that $r$
takes a universal, $d$--independent value. The constant volume property
is further studied by computing $L_n^2 = \left(\prod_{i=1}^6
\vev{\lambda_i}_{n,V}\right)^{1/6}$. The results are shown in the
right plot of Fig. \ref{f:01}. For $0.5<x<1$ $L_n^2 \approx \ell
\approx 0.627$ as predicted by the GEM.

Finally we attempt to compare the free energies ${\cal F}_3(x)$ and
${\cal F}_4(x)$ by numerically computing $\Delta_{34}=-{\cal
  F}_3(x)+{\cal F}_4(x)$. Their value at the solution of Eq. \rf{2.10}
compares the free energies of the SO(3) and SO(2) vacua
respectively. The calculation is done as described in \cite{multifac}
by computing
\begin{equation}
\label{3.2}
 \Delta_{34} = \frac{1}{N^2} \log w_3 (x_s) - \frac{1}{N^2} \log w_4 (x_s) - \int^{\textrm{SO(3)}}_{\textrm{SO(2)}} \frac{1}{N^2} \frac{d}{dx} \log \rho^{(0)} (x) 
 \simeq \frac{1}{N^2} \log w_3 (x_s) - \frac{1}{N^2} \log w_4 (x_s) 
\end{equation}
at $x_s\approx 0.31$. The term 
$\displaystyle\int^{\textrm{SO(3)}}_{\textrm{SO(2)}}  
\frac{1}{N^2}\frac{d}{dx}\log \rho^{(0)}(x)$ 
vanishes  in the large $N$ limit. The comparison of the free
energies can be read off
the right plot of Fig. \ref{f:03} where the $N=24$ results
are shown. Although we cannot draw a definite conclusion, we see no
inconsistency with a dominating SO(3) vacuum as predicted by GEM.

In summary, we simulated from first principles a 6 dimensional version
of the IIB matrix model. We studied the scenario of dynamical
compactification by SSB of the eigenvalues of the tensor $T_{\mu\nu}$
and found that our results are consistent with the quantitative
predictions of the GEM. The importance of the fluctuating phase in
inducing SSB was demonstrated first by confirming that no SSB occurs
in the phase quenched model. A strong complex action problem was found
to hinder ordinary Monte Carlo simulations and the factorization
method was used in order to determine and sample the region favored
by the competing effects of the fluctuating phase, the real part of
the action and the density of states. The distribution function
$\rho_n(x)$ of the eigenvalues $\vev{\tilde\lambda_n}$ were considered
and by using the scaling properties of the factors $f_n^{(0)}(x)$ and
$w_n(x)$ we computed the scale $r$ of the compactified
dimensions. The results were found to be consistent with the unique
$d$--independent numerical value computed using the GEM and with the
constant volume property $R^{d} r^{6-d} = \ell^6$. A strong, non
perturbative, hard core potential against the collapse of the
eigenvalues is shown to be generated dynamically which plays a crucial
role in obtaining non trivial solutions in the large $N$ limit,
generating scales $r$ and $R$ that are comparable and giving finite
spacetime volume.

Recent results in \cite{Kim:2011cr}, however, have shown that
it is possible to study the {\it Lorentzian} IIB matrix model and
obtain an expanding, large, 3 dimensional space which arises after a
critical time\footnote{In fact, it turned out that the complex action problem
can be totally avoided in the Lorentzian model in a highly nontrivial
manner \cite{Kim:2011cr}.}. 
This is a very exciting possibility, which
motivates further study of the IIB matrix model as a non perturbative
definition of superstring theory and as a model for string inspired
cosmology.


\end{document}